# Co-evaporated Formamidinium tin triiodide with suppressed p-type self-doping

*Junhyoung Park[1], Andrea Olivati[1], Mirko Prato[2], Min Kim[3], Annamaria Petrozza[1]*

[1] Center for Nano Science and Technology@Polimi, Istituto Italiano di Tecnologia, via Rubattino 81, Milano 20134, Italy

[2] Materials Characterization Facility, Istituto Italiano di Tecnologia, Via Morego, Genova, 16163 Italy

[3]Department of Chemical Engineering, University of Seoul, 163, Seoulsiripdae-ro, Dongdaemun-gu, Seoul, 02504, South Korea

**Abstract**

Co-evaporation of formamidinium tin triiodide (FASnI$_3$) precursors, without any additives or reducing agents, leads to the growth of a highly crystalline thin film which shows a bandgap around 1.31 eV, closely matching the theoretical value predicted from the ideal single crystal structure of FASnI$_3$. The polycrystalline thin film presents a lower tendency of Sn$^{2+}$ to Sn$^{4+}$ oxidation and highly reduced tendency to self-doping, demonstrating, overall, an improved resistance to defects formation. These findings suggest solvent-free co-evaporation processes as a promising route for high quality Sn-based perovskite polycrystalline thin films.

Metal halide perovskites are revolutionizing optoelectronic applications, including solar cells, LED, and photodetectors due to their exceptional optical properties. Tin halide perovskites (THPs) are regarded as promising candidates for applications like solar cells, near-infrared LEDs, and photodetectors due to their narrower bandgaps compared to lead halide perovskites (LHPs). However, chemically active lone pair electrons and low Sn-vacancy ($V_{Sn}$) formation energy promote $Sn^{2+}$ to $Sn^{4+}$ oxidation. This causes self-doping of THPs which exhibit p-type semiconductor characteristics. While, *per se,* it is a feature which may be exploited in different (opto)electronic and photonic applications, the self-doping is currently not under control, which makes THPs semiconductor not a reliable class of materials. So far, solution-processed THPs thin film development was primarily conducted via additive engineering with the attempt of controlling defects formation. With all efforts, THPs still suffer from chemical instability due to inevitable solvent-induced defects and remnant molecules. Consequently, solvent-free evaporation techniques, also offering precise stoichiometry control, large-area uniformity, and processing in $O_2$/moisture-free environment, are especially advantageous for THPs fabrication, enabling constraint of solvent related issues.

However, only a few attempts have been shown in this direction, without a detailed characterization of the thin film properties. Specifically, co-evaporated methylammonium tin triiodide ($MASnI_3$) has shown poor morphology and optical characteristic than spin-coated one.[1] Although it is not clear, this might be due to the diffusive nature of MAI evaporation and the inherent instability of $MA^+$. In contrast, Formamidinium ($FA^+$)-based perovskites generally has exhibited higher thermal stability and performance compared to $MA^+$-based ones, attributable to lower volatility and reduced fragmentation of $FA^+$. Despite this, the successful co-evaporation of $FASnI_3$ and investigation on its properties remain unreported.

Here, we report for the first highly crystalline, additive-free, co-evaporated pure $FASnI_3$ film. We compare stoichiometric co-evaporated and spin-coated $FASnI_3$ (with additive) films to highlight their distinct material and optical properties. Co-evaporated films exhibited a pin-hole free flat surface with compact morphology compared to spin-coated ones. X-ray diffraction (XRD) patterns (Figure 1a) show co-evaporated $FASnI_3$ peaks closely match simulated positions for ideal cubic (Pm3m), unlike significant deviations in spin-coated peaks.[2] (Figure S1). The shift of XRD peaks toward higher angles and reduced full width at half maximum (FWHM) in co-evaporated films imply they experiences reduced non-uniform/uniform lattice strain, likely due to more homogeneous crystallization on the substrate. [3] In contrast, the spin-coated film may experience greater structural disorder and tensile strain. [4]

Optical characterizations further support this structural distinction. The spin-coated film exhibits a blue-shifted absorption onset and high background (~850 nm, Fig 1b), suggesting high doping levels and broader distribution of sub-bandgap states.[5] In contrast, the co-evaporated film shows a sharper absorption edge, implying well-defined density of states near the edge. Consistently, PL spectrum of co-evaporated film shows minimal blue-shift with narrow FWHM (37.2 nm), suggesting negligible Moss-Burstein effect, indicative of reduced electronic doping while spin-coated one show broad and blue-shifted PL spectrum. [6] A Tauc plot analysis further reveals the co-evaporated film has a narrower optical bandgap (~1.31 eV) compared to previously reported state-of-the-art spin-coated films (~1.40 eV, Figure 1c). [7] This result is also consistent with DFT calculations (HSE06-D3) for $FASnI_3$ bandgap. [8] From combined understanding, we propose that the tensile strain in spin-coated films reduces the overlap between Sn 5s and I 5p orbitals, thereby increasing the energy gap between bonding and antibonding states.

XPS analysis reveals distinct surface chemical characteristics between films as shown in Figure 1d. The co-evaporated film predominantly showed $Sn^{2+}$ signals (Sn $3d_{5/2}$ BE: 486.4 ± 0.2 eV), while spin-coated $FASnI_3$ exhibited significant $Sn^{4+}$ states (487.5 eV), consistent with prior reports (Figure 1d). Instead, a minor component, which wasn't clearly revealed, appeared at 487.1 eV in the co-evaporated film. [9] For I $3d_{5/2}$, co-evaporated films displayed a single peak (619.2 eV), whereas spin-coated films showed double peaks (619.2 eV and 620.2 eV), suggesting rapid surface degradation even under strictly controlled conditions ($O_2/H_2O$ < 0.5 ppm). This suppressed $Sn^{4+}$ formation consistently indicates minimal Sn-vacancy formation in co-evaporated films, reducing charge-compensating defects and background holes. UPS results (Figure S2) corroborate this, showing a mid-gap Fermi level, confirming close to intrinsic semiconductor characteristics of co-evaporated films.

Further, to show p-doping controllability of co-evaporation, we investigated material/optical property dependence on FAI partial pressure ($P^{FAI}$). Recognizing that precursor interactions strongly affect sticking coefficients [10], we maintained base pressure near $5×10^{-7}$ mTorr while precisely controlling the FAI evaporation rate (± 0.01 Å/s) using a dedicated quartz crystal monitors. Upon initiating evaporation, the total pressure briefly rose to approximately $6×10^{-6}$ mTorr before stabilizing. Four distinct deposition rate ratios ($FAI/SnI_2$) were nominally defined– $P^{(FAI0.5)}$, $P^{(FAI0.75)}$, $P^{(FAI1.0)}$, and $P^{(FAI1.5)}$–reflecting $P^{FAI}$. (Figure S3) $P^{FAI}$ markedly influences microstructural, crystallographic, optical, and electrical properties. While all conditions produced highly absorptive films, substantial differences in microstructure and crystallinity emerged (Figure S4 and S5). Conditions $P^{(FAI0.75)}$ and $P^{(FAI1.0)}$ yield highly crystalline and uniformly structured films. In contrast, $P^{(FAI0.5)}$ and $P^{(FAI1.5)}$ led to amorphous phases without long-range order.

Optical analysis further elucidates the structural quality dependence on $P^{\text{FAI}}$ (Figure 1e). Films deposited under $P^{(\text{FAI1.0})}$ exhibit the steepest Urbach tail, indicative of superior crystallinity, whereas flatter Urbach tails observed for $P^{(\text{FAI0.5})}$ and $P^{(\text{FAI1.5})}$ signified energetic disorder, consistent with structural analyses. Correspondingly, PL spectra (Figure 1f) indicate emission peaks at longer wavelengths (~950 nm, 1.31 eV) specifically for $P^{(\text{FAI1.0})}$, affirming minimal Moss-Burstein effect from a reduced self-doping. Fluorescence-dependent PLQY measurements elucidate defects and doping effects. $P^{(\text{FAI0.5})}$ and $P^{(\text{FAI0.75})}$ showed lower relative PLQY due to dominant nonradiative recombination. (Figure 1g) $P^{(\text{FAI1.0})}$ exhibited the highest relative PLQY. Interestingly, all samples show a power dependance of the PLQY, though with different slopes, until Auger-like process kick-in. This is well in agreement with low electronic doping densities. Accordingly, resistivity measurements (Figure 1h) also explain the effect of doping. Films deposited at $P^{(\text{FAI0.5})}$, $P^{(\text{FAI0.75})}$, $P^{(\text{FAI1.0})}$, and $P^{(\text{FAI1.5})}$ exhibited resistivities of approximately 2.57 $k\Omega{\cdot}m$, 7.4 $\Omega{\cdot}m$, 13.3 $\Omega{\cdot}m$, and 42.5 $\Omega{\cdot}m$, respectively, substantially higher than those typically observed in spin-coated FASnI$_3$ (without additive: $10^{-2}$-$10^{-4}$ $\Omega{\cdot}m$ [8] and with 10 % of SnF$_2$ $1.3*10^{-2}$ $\Omega{\cdot}m$ [11]). Higher resistivity indicates decreased hole concentration from self-doping.

This systemic research demonstrates co-evaporation–when performed without additives–can produce an FASnI$_3$ film with restrained p-doping. Co-evaporated films exhibit ideal optical and structural properties, closely matching computational predictions, highlighting the significant influence of co-evaporation on THP quality. It deepens understanding of FASnI$_3$ co-evaporation, establishing it as a promising route for high-quality THP and offering a fundamental framework for extending into optoelectronic applications.

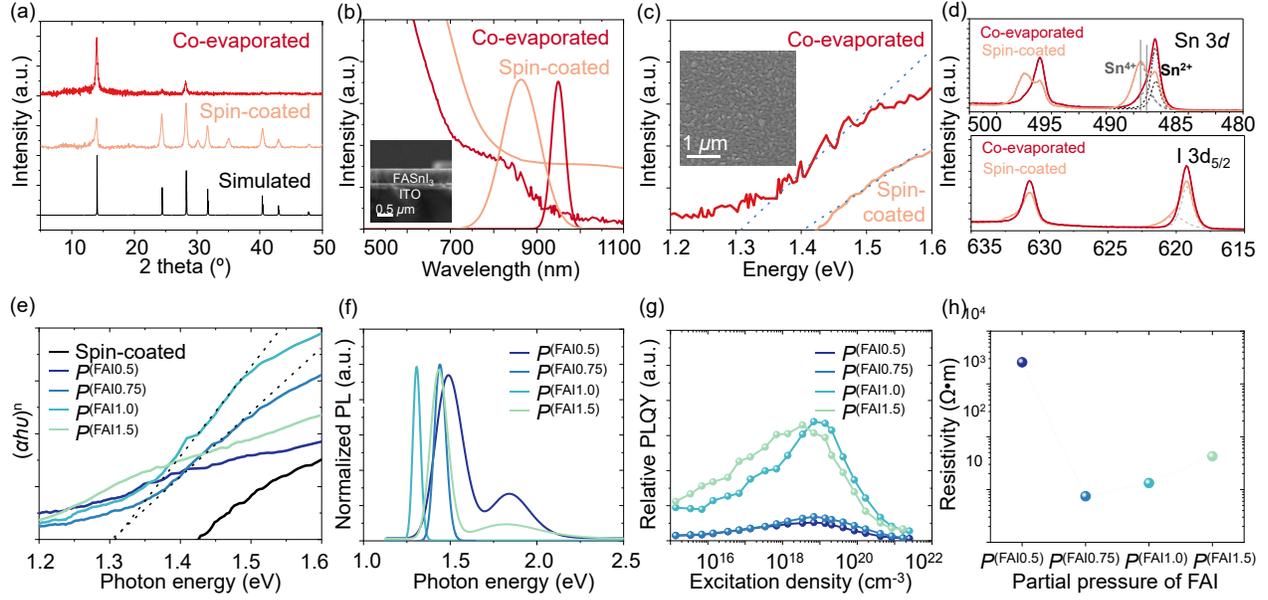

**Figure 1.** (a) X-ray diffraction (XRD) patterns comparing three cases: an ideal cubic structure calculated computationally, an experimentally spin-coated, and co-evaporated FASnI₃ film. (b) Optical characterization through absorption and photoluminescence spectra of co-evaporated and spin-coated FASnI₃ films. Cross-sectional SEM image of the co-evaporated FASnI₃ film (~400 nm thickness), demonstrating a highly uniform and smooth surface. (c) bandgaps estimated from Tauc method of Spin-coated and Co-evaporated FASnI₃ films. (d) X-ray photoelectron spectroscopy (XPS) spectra showing the binding energies of Sn $3d_{5/2}$ and I $3d_{5/2}$ states in co-evaporated FASnI₃. (e) Tauc plot analysis of co-evaporated FASnI₃ films prepared at four distinct partial pressures of Formamidinium iodide (FAI), denoted as $P^{(FAI0.5)}$, $P^{(FAI0.75)}$, $P^{(FAI1.0)}$, and $P^{(FAI1.5)}$, listed in order of increasing partial pressure. (f) Photoluminescence spectra plotted against photon energy for co-evaporated FASnI₃ films fabricated under conditions $P^{(FAI0.5)}$, $P^{(FAI0.75)}$, $P^{(FAI1.0)}$, and $P^{(FAI1.5)}$. (g) Excitation density dependent Photoluminescence Quantum Yield (PLQY) of co-evaporated FASnI₃ films fabricated under conditions $P^{(FAI0.5)}$, $P^{(FAI0.75)}$, $P^{(FAI1.0)}$, and $P^{(FAI1.5)}$. (h) Resistivity of co-evaporated FASnI₃ films corresponding to conditions $P^{(FAI0.5)}$, $P^{(FAI0.75)}$, $P^{(FAI1.0)}$, and $P^{(FAI1.5)}$.

# Supporting Information

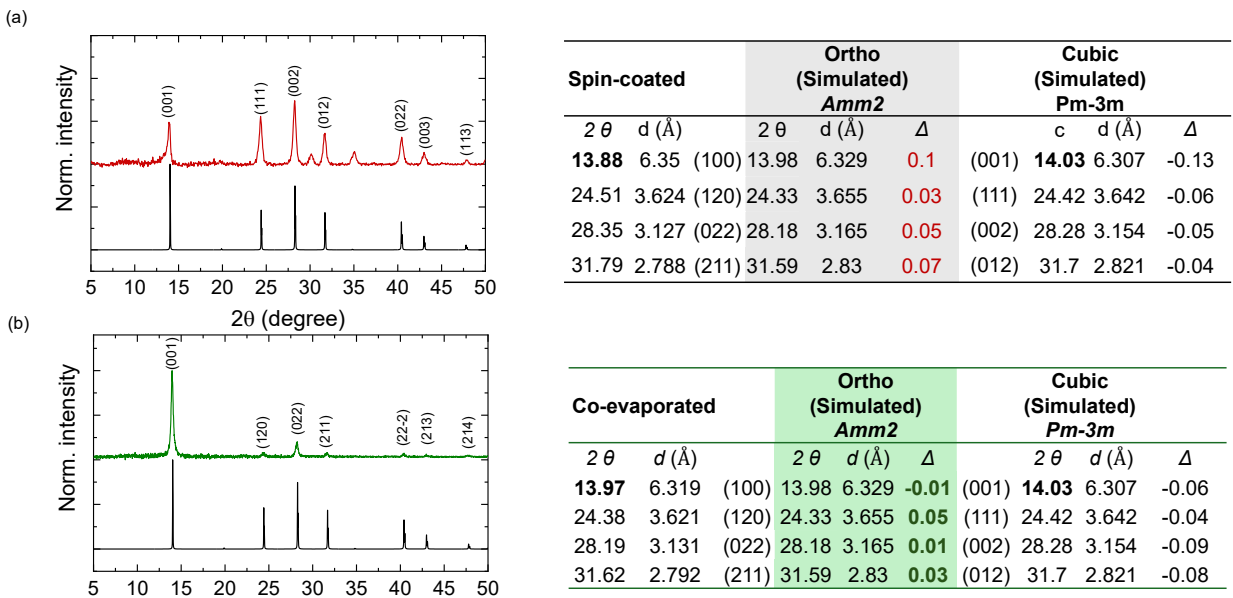

| Spin-coated | | Ortho (Simulated) *Amm2* | | | Cubic (Simulated) *Pm-3m* | | |
|---|---|---|---|---|---|---|---|
| *2θ* | d (Å) | *2θ* | d (Å) | *Δ* | c | d (Å) | *Δ* |
| **13.88** | 6.35 (100) | 13.98 | 6.329 | 0.1 | (001) | **14.03** | 6.307 | -0.13 |
| 24.51 | 3.624 (120) | 24.33 | 3.655 | 0.03 | (111) | 24.42 | 3.642 | -0.06 |
| 28.35 | 3.127 (022) | 28.18 | 3.165 | 0.05 | (002) | 28.28 | 3.154 | -0.05 |
| 31.79 | 2.788 (211) | 31.59 | 2.83 | 0.07 | (012) | 31.7 | 2.821 | -0.04 |

| Co-evaporated | | Ortho (Simulated) *Amm2* | | | Cubic (Simulated) *Pm-3m* | | |
|---|---|---|---|---|---|---|---|
| *2θ* | d (Å) | *2θ* | d (Å) | *Δ* | *2θ* | d (Å) | *Δ* |
| **13.97** | 6.319 (100) | 13.98 | 6.329 | -0.01 | (001) | **14.03** | 6.307 | -0.06 |
| 24.38 | 3.621 (120) | 24.33 | 3.655 | 0.05 | (111) | 24.42 | 3.642 | -0.04 |
| 28.19 | 3.131 (022) | 28.18 | 3.165 | 0.01 | (002) | 28.28 | 3.154 | -0.09 |
| 31.62 | 2.792 (211) | 31.59 | 2.83 | 0.03 | (012) | 31.7 | 2.821 | -0.08 |

Figure S1. (a) X-ray diffraction patterns of spin-coated pure FASnI₃. The peak position gap between computationally calculated ideal orthorhombic (*Amm2*) and cubic (*Pm-3m*) are huge. (b) X-ray diffraction patterns of co-evaporated FASnI₃ which are close to ideal structure calculated from ideal orthorhombic (*Amm2*) and cubic (*Pm-3m*).

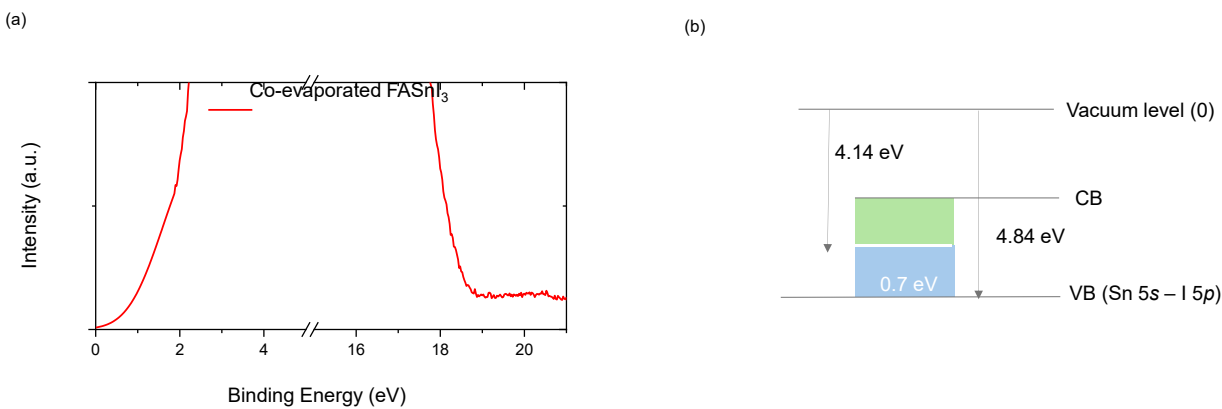

Figure S2. (a) Ultra-violet photoelectron spectroscopy measured on co-evaporated FASnI₃. (b) Estimated conduction band minima and Valence band maxima energy estimated from UPS

measurements. The correct fittings of the spectra were conducted using Casa XPS software and built-in algorism for fitting.

| | Total pressure (mTorr) |
|---|---|
| $P^{(FAI1.5)}$ | $4\sim6 \times 10^{-6}$ |
| $P^{(FAI1.0)}$ | $1.6\sim2.5 \times 10^{-6}$ |
| $P^{(FAI0.75)}$ | $1.2\sim1.4 \times 10^{-6}$ |
| $P^{(FAI0.5)}$ | $0.9\sim1.2 \times 10^{-6}$ |

Figure S3. Total pressure was tracked using PKR 251 absolute pressure sensor (Pirani/Penning) horizontally installed approximately same height with the substrate in the evaporator. The pressure of the vacuum was Pfeiffer Display control unit 110.

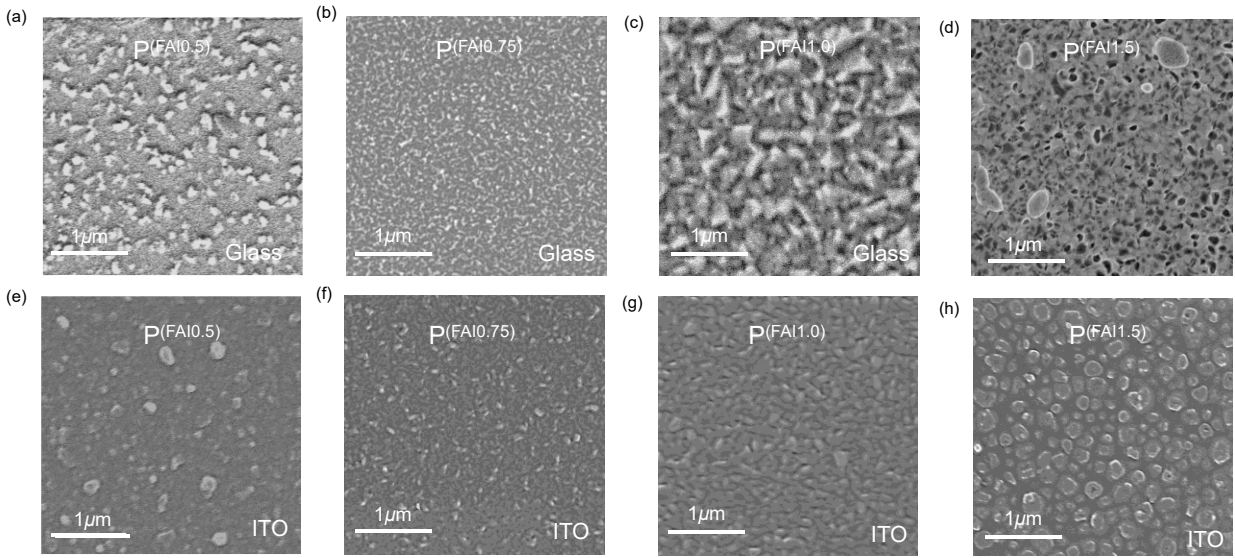

Figure S4. (a) FASnI$_3$ film fabricated on glass under $P^{(FAI0.5)}$ condition. (b) under $P^{(FAI0.75)}$ condition. (c) $P^{(FAI1.0)}$ condition (d) $P^{(FAI1.5)}$ condition. (e) FASnI$_3$ film fabricated on ITO glass under $P^{(FAI0.5)}$ condition. (f) FASnI$_3$ under $P^{(FAI0.75)}$ condition. (g) under $P^{(FAI1.0)}$ condition. (h) under $P^{(FAI1.5)}$ condition.

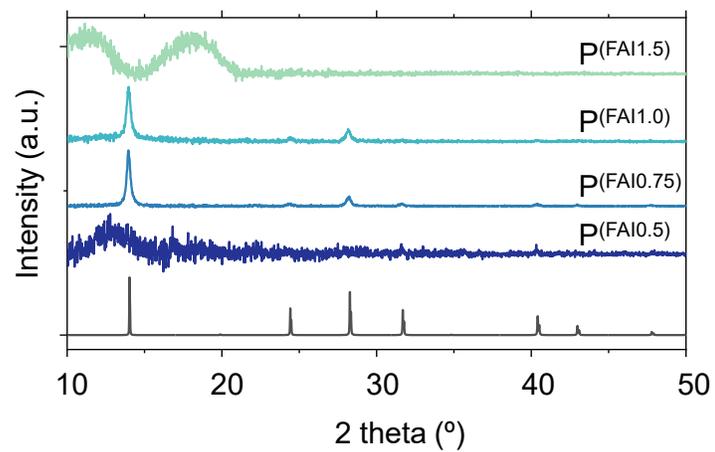

Figure S5. X-ray diffraction patterns were obtained from co-evaporated FASnI$_3$ films. These films were prepared under four distinct partial pressures of Formamidinium iodide (FAI), denoted in increasing order of partial pressure as $P^{(FAI0.5)}$, $P^{(FAI0.75)}$, $P^{(FAI1.0)}$, and $P^{(FAI1.5)}$

# AUTHOR INFORMATION

Corresponding Author

Annamaria Petrozza - [1]Center for Nano Science and Technology@Polimi, Istituto Italiano di Tecnologia, via Rubattino 81, Milano 20134, Italy; ORCID: http://orcid.org/0000-0001-6914-4537; Email: annamaria.petrozza@iit.it

Junhyoung Park - [1]Center for Nano Science and Technology@Polimi, Istituto Italiano di Tecnologia, via Rubattino 81, Milano 20134, Italy; ORCID: https://orcid.org/0000-0001-6078-5004-5004; Email: jhpark0130@gmail.com

# AUTHOR CONTRIBUTION

J. Park (Fisrt author, Corresponding author): Conceptualization, Methodology, Data Curation, Formal analysis, Writing -Original draft, Writing-Reviewing and editing), A. Petrozza (Corresponding author): Conceptualization, Writing-Reviewing and editing, Supervision, and Funding acquisition, A. Olivati (Second author): Formal analysis, M. Prato (Third author): Formal analysis, Proof reading, M.Kim (Fourth author): Scientific discussion and Writing-Reviewing and editing.

## Funding Sources

HORIZON Grants VALHALLA (no. 101082176)

# ACKNOWLEDGMENT

The authors acknowledge the support from the projects supported by the HORIZON Grants VALHALLA (no. 101082176). This author appreciates Prof. Lorenzo Malavasi, Isabella Poli and Giulia Folpini for educational discussion.

# ABBREVIATIONS

THP, Tin halide perovskite; LHP, Lead halide perovskite; MASnI$_3$, Formamidinium tin triiodide;

FASnI$_3$, Formamidinium tin triiodide; FAI, Formamidinium iodide; MAI, Formamidinium

iodide; PLQY, Photoluminescence quantum yield

**Table of Contents**

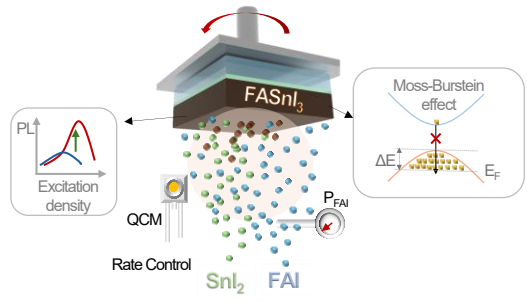